\documentclass[aps,prd,epsf]{revtex4}
\usepackage[latin1]{inputenc}
\usepackage[english]{babel}
\usepackage{latexsym}
\usepackage{amssymb}
\usepackage{graphicx}
\newcommand{\be}{\begin{equation}}
\newcommand{\ee}{\end{equation}}
\newcommand{\bea}{\begin{eqnarray}}
\newcommand{\eea}{\end{eqnarray}}

\def\PR{P_{r}}
\def\PM{P_{m}}
\def\PD{P_{d}}
\def\ty_3{\tilde{y}_3}

\begin{document}

%____________________________________________________________________________
\title{$f(R)$ gravity theories in Palatini formalism: \\
cosmological dynamics and observational constraints}

%______________________________________                                                            

\author{Stéphane Fay$^{1,2}$\footnote{steph.fay@gmail.com} and
Reza Tavakol$^1$\footnote{r.tavakol@qmul.ac.uk}}

\affiliation{$^1$School of Mathematical Sciences,\\
Queen Mary, University of London,
London E1 4NS, UK}
\affiliation{$^2$Laboratoire Univers et Théories (LUTH), UMR 8102\\
Observatoire de Paris, F-92195 Meudon Cedex, France}

\author{Shinji Tsujikawa\footnote{shinji@nat.gunma-ct.ac.jp}}
\affiliation{Department of Physics, Gunma National College of
Technology, Gunma 371-8530, Japan}

%______________________________________
\begin{abstract}
%---------------------------------------
    
We make a systematic study of the cosmological dynamics 
for a number of $f(R)$ gravity theories in Palatini formalism,
using phase space analysis as well as numerical simulations.
Considering homogeneous and isotropic models, 
we find a number of interesting results:
(i) models based on theories of the type (a) $f(R)=R-\beta/R^n$
and (b) $f(R)=R+\alpha {\rm ln}\, R-\beta$, unlike the 
metric formalism, are capable of producing the 
sequence of radiation-dominated, matter-dominated and 
de-Sitter periods, and (ii) models based on theories of the type
(c) $f(R)=R+\alpha R^m-\beta/R^n$
can produce early as well as late accelerating phases
but an early inflationary epoch does not seem to be 
compatible with the presence of a subsequent radiation-dominated era. 
Thus for the classes of models considered here,
we have been unable to find the sequence of all
four dynamical epochs required to account for 
the complete cosmological dynamics, even though three out 
of four phases are possible.

We also place observational constraints on these models 
using the recently released supernovae data by the Supernova 
Legacy Survey as well as 
the baryon acoustic oscillation peak in the SDSS 
luminous red galaxy sample and the CMB shift parameter.
The best-fit values are found to be $n=0.027, \alpha=4.63$
for the models based on (a) and $\alpha=0.11, \beta=4.62$
for the models based on (b),
neither of which are significantly preferred over the 
$\Lambda$CDM model. 
Moreover, the logarithmic term alone is not capable
of explaining the late acceleration.
The models based on (c) are also consistent with the data 
with suitable choices of their parameters.

We also find that some of the models for which the 
radiation-dominated epoch is absent prior to the matter-dominated 
era also fit the data. 
The reason for this apparent contradiction is that 
the combination of the data considered here does not 
place stringent enough constraints on the cosmological 
evolution prior to the decoupling epoch, 
which highlights the importance
of our combined theoretical-observational approach 
to constrain models.

\end{abstract}

%-----------------------------------%
\maketitle

%----------------------------------
\section{Introduction} \label{s0}
%----------------------------------

Recent high--precision observations by the Wilkinson Microwave
Anisotropy Probe (WMAP), together with other Cosmic Microwave
Background (CMB) and high redshift surveys have produced a wealth of
information regarding the early universe. The analysis of the resulting data
has provided strong evidence for the core predictions of 
the inflationary cosmology, 
including the almost spatial flatness of the universe \cite{Spe03,Spe06}.
Furthermore, these observations coupled with the low redshift 
supernovae surveys \cite{Rie98,Per99,DalDjo03,Rie04,Ast06}
and observations of large scale structure
\cite{Tegmark-etal04,Seljak-etal05} and baryon acoustic 
oscillations \cite{Eis05} suggest that the universe is at
present undergoing a phase of accelerated expansion
(see Refs.~\cite{review,review2} for reviews).
Thus a `standard' model of cosmology has emerged which
is characterised by four distinct phases:
accelerated expansions at both early 
and late times, mediated by radiation-dominated and matter-dominated 
eras. The central question in cosmology at present is, 
therefore, how to successfully account for these
distinct dynamical phases in the history of the universe,
and in particular whether they can all be realised 
simultaneously within a single theoretical framework,
motivated by a fundamental theory of 
quantum gravity. 

A number of scenarios have been proposed to account
for these dynamical modes of behaviour. These fall into two categories:
(i) those involving the introduction of exotic matter sources, 
and (ii) those introducing changes to the gravitational sector of 
General Relativity. Among the latter are $f(R)$ gravity theories, 
which involve non-linear generalisations to the (linear) Hilbert action. 
Nonlinear modifications are expected to be present
in the effective action of the gravitational field when string/M-theory
corrections are considered 
\cite{Buchbinder-et-al-1992,Gasperini-Venezinano-1992,Noriji-etal-2003,Vassilevich-2003}.

A great deal of effort has recently gone into the study of such theories.
An important reason for this interest 
has been the demonstration that generalised Lagrangians of this type -
which include negative as well as positive powers of the curvature
scalar - can lead to accelerating phases both at early \cite{Sta}
and late \cite{Carroll03,Capo03} 
times in the history of the universe
(see also Ref.~\cite{CapoLuca}).

In deriving the Einstein field equations from the Hilbert action 
the variations are taken
with respect to the metric coefficients, while the connections are
assumed to be the Christoffel symbols  defined in terms of the metric.
An alternative procedure - the so called Palatini approach,
originally considered by Einstein himself  - is
to treat both the metric and the (affine) connections as independent  
variables and perform the variations with respect to both.
In the case of linear Hilbert action 
both approaches produce identical results,
as long as the energy-momentum tensor does not depend on the 
connection \cite{Wald}.
This, however, is not the case once the gravitational
Lagrangian is allowed to be nonlinear.
In that case the two methods of variation produce different 
field equations with nontrivial differences in the resulting dynamics.

Performing variations of nonlinear actions using the metric approach 
results in field equations that are fourth order;
which makes them difficult to deal with in practice. 
Furthermore, within this framework,
models based on theories of the type $f(R)=R-\beta/R^n$
have difficulties passing the solar system 
tests \cite{Chiba03} and having the correct Newtonian 
limit \cite{Sot}. In addition such theories
suffer gravitational instabilities as discussed in Ref.~\cite{DK03}.
Also recent studies have found that these theories are
not able to produce a standard matter-dominated era followed by
an accelerated expansion~\cite{APT,AGPT}.
Finally, models based on theories of the type 
$f(R)=R+\alpha R^m-\beta/R^n$ 
have been shown to have difficulties in satisfying the set of constraints
coming from early and late-time acceleration, Big Bang 
Nucleosynthesis and fifth-force experiments \cite{BBH}.
(See Refs.~\cite{more} for other works concerning the metric 
formalism.)

Variations using the Palatini approach \cite{Vollick,Fla}, 
however, result in second order field equations which are in addition
known to be free of instabilities \cite{Meng03,Meng04}.
Such theories have also been shown to satisfy
the solar system tests and have the correct Newtonian 
limit \cite{Sotnewton}.
Here we shall concentrate on the Palatini approach and 
consider a number of families of $f(R)$ theories recently put forward
in the literature. These theories have been the focus of 
great deal of interest recently with a number 
of studies attempting to determine their viability 
as cosmological theories, both
theoretically \cite{Sot,NOlnR,MenglnR,Sot2,Palapapers} and 
observationally \cite{AmaElg06}.
Despite these efforts, it is still fair to say that 
cosmological dynamics of models based on such theories 
is not fully understood. Of particular interest is
the number of dynamical phases such models are capable of admitting, 
among the four phases required
for strict cosmological viability, namely
early and late accelerating phases mediated
by radiation-dominated and matter-dominated epochs.
Especially it is important to determine whether
they are capable of allowing all four phases required
for cosmological evolution. We should note that the
idea that such theories should be capable of
successfully accounting for all four phases 
is clearly a maximal demand. It would still be of great interest 
if such theories could successfully account for a sequence 
of such phases, such as the first or the last
three phases. For example the presence of the last 3 phases could 
distinguish these from the corresponding
theories based on the metric formalism \cite{APT}.
Also observational constraints have so far only been obtained
for the models based on the theories of the type $f(R)=R-\beta/R^n$.
Such constraints need to be also studied for other theories considered 
in this context in the literature.

Here in order to determine the cosmological viability of models based
on $f(R)$ theories in Palatini formalism we shall employ 
a two pronged approach.
Considering homogeneous and isotropic
settings, we shall first provide autonomous equations applicable
to any $f(R)$ gravity theory.
We shall then make a systematic and detailed
phase space analysis of a number of families of $f(R)$ theories.
This provides a clear understanding of the dynamical modes 
of behaviour which are admitted by these theories, 
and in particular whether 
these theories possess early and late accelerating phases which are
mediated by the radiation-dominated and matter-dominated 
eras respectively.
The existence of these phases is a necessary but not sufficient
condition for cosmological viability of such theories.
To be cosmologically viable, it is also necessary
that the subset of parameters in these theories that allows these
phases to exist are also compatible with 
observations \footnote{The idea that such theories should be able to
successfully account for all four phases in the cosmological evolution
is clearly a maximal demand. It would still be of great interest
if such theories could successfully account for a sequence
of such phases including an accelerating phase, 
such as the first or the last
three phases.}.
We constrain these parameters for a number of families of 
$f(R)$ theories using the data from recent observations,
including recently released supernovae data
by the Supernova Legacy Survey \cite{Ast06} 
as well as the baryon acoustic oscillation peak in the 
Sloan Digital Sky Survey (SDSS) luminous red galaxy
sample \cite{Eis05}
and the CMB shift parameter \cite{Spe03,Spe06}.

The plan of the paper is as follows. 
In Section \ref{s2} we give a brief
account of the Palatini formalism and provide 
the basic equations for general $f(R)$ theories.
We also introduce new variables which 
allow these equations to be written as autonomous dynamical systems.
In Section \ref{s3} we proceed to study the cosmological dynamics 
for a number of classes of $f(R)$ theories.
Particular emphasis will be placed on 
finding cases which can allow the largest number
of dynamical phases required in the cosmological evolution.
In Section \ref{s4} we obtain observational 
constraints on these theories. 
This allows further constraints to be placed on the parameters
of these theories.
Finally we conclude in Section \ref{s5}.

%-------------------------------------------------------------------
\section{$f(R)$ theories in Palatini formalism and autonomous 
equations} 
\label{s2}
%-------------------------------------------------------------------

We consider the classes of theories given by the generalised action
\begin{eqnarray}
\label{action}
S=\int{\rm{d}}^{4}x\sqrt{-g}
\left[\frac{1}{2\kappa}f(R)+{\cal{L}}_{m}
+{\cal{L}}_r \right]\,,
\end{eqnarray}
where $f$ is a differentiable function of
the Ricci scalar $R$, ${\cal L}_m$ and  ${\cal L}_r$ are
the Lagrangians of the pressureless dust and
radiation respectively, $\kappa=8\pi G$ and 
$G$ is the gravitational constant.
Motivated by recent observations we shall study the cosmological dynamics in
these theories for a flat Friedmann-Lemaître-Robertson-Walker (FLRW) background
\begin{eqnarray}
\label{flrw}
{\rm d}s^2=-{\rm d}t^2+a^2(t) {\rm d}{\bf x}^2\,,
\end{eqnarray}
where $a(t)$ is the scale factor and $t$ is the cosmic time.

As was mentioned above, in Palatini formalism the
metric and the affine connections are treated as independent
variables with respect to which the action is varied.
Varying the action (\ref{action}) with respect to the 
metric $g_{\mu \nu}$ gives 
(see e.g. Ref.~\cite{Sot})
%------------------------------------%
\begin{equation}
\label{eq1}
FR_{\mu\nu}-\frac12 fg_{\mu\nu}=
\kappa T_{\mu\nu}\,,
\end{equation}
%------------------------------------%
where $F \equiv \partial f/\partial R$ 
and $T_{\mu\nu}$ is given by
\begin{equation}
T_{\mu\nu} = -\frac{2}{\sqrt{-g}}
\frac{\delta [{\cal L}_m + {\cal L}_r]}{\delta g^{\mu\nu}}\,.
\end{equation}
For the FLRW metric with pressureless dust and radiation we obtain the 
generalised Friedmann equation 
\begin{eqnarray}
\label{baeq1}
6F \left( H+\frac{\dot{F}}{2F} \right)^2-f
=\kappa (\rho_m+2\rho_r)\,,
\end{eqnarray}
where a dot denotes differentiation with respect to 
$t$ and $H \equiv \dot{a}/a$ is the Hubble parameter. 
Here $\rho_m$ and $\rho_r$ are the energy densities 
of matter and radiation, respectively, which satisfy
\begin{eqnarray}
\label{rhom}
& & \dot{\rho}_m+3H \rho_m=0\,, \\
& & \dot{\rho}_r+4H \rho_r=0\,.
\end{eqnarray}
Contracting Eq.~(\ref{eq1}), and recalling that the trace of
the radiative fluid vanishes, gives
%------------------------------------%
\begin{equation}\label{eq3}
FR-2f= -\kappa \rho_m \,.
\end{equation}
%------------------------------------%
Using Eqs.~(\ref{baeq1})-(\ref{eq3}) we obtain
\begin{equation}
\label{baeq3}
\dot{R}=\frac{3\kappa H \rho_{m}}
{F'R-F}=-3H \frac{FR-2f}{F'R-F}\,,
\end{equation}
where a prime denotes a derivative with respect to $R$.
Combining Eqs.~(\ref{baeq1}) and (\ref{baeq3}) we find
\begin{eqnarray}
\label{Hubble}
H^2=\frac{2\kappa (\rho_{m}+\rho_r)+FR-f}
{6F\xi}\,,
\end{eqnarray}
where 
\begin{eqnarray}
\xi \equiv \left[ 1-\frac32 \frac{F' (FR-2f)}
{F (F'R-F)} \right]^2\,.
\end{eqnarray}
In the case of the Hilbert action with $f=R$, Eq.~(\ref{Hubble}) reduces to the
standard Friedmann equation: $H^2=\kappa (\rho_m+\rho_r)/3$.

To obtain a clear picture of possible dynamical regimes
admitted by these theories, we shall
%in this Section, 
make a detailed study of a number of families of $f(R)$ theories
of the type (\ref{action}), using phase space analysis.
It is convenient to express these systems as
autonomous systems by introducing the following 
dimensionless variables:
\begin{eqnarray}
y_1 \equiv \frac{FR-f}{6F\xi H^2}\,, \quad
y_2 \equiv \frac{\kappa \rho_r}{3F \xi H^2}\,.
\label{y1y2}
\end{eqnarray}
In terms of these variables the constraint equation 
(\ref{Hubble}) becomes
\begin{eqnarray}
\label{constraint}
\frac{\kappa \rho_{m}}{3F \xi H^2}=1-y_1-y_2\,.
\end{eqnarray}
Differentiating Eq.~(\ref{Hubble}) and using 
Eqs.~(\ref{baeq1})-(\ref{Hubble}) we obtain
\begin{eqnarray}
\label{Hdot}
2\frac{\dot{H}}{H^2}=-3+3y_1-y_2-\frac{\dot{F}}{HF}
-\frac{\dot{\xi}}{H\xi}+\frac{\dot{F}R}{6F \xi H^3}\,.
\end{eqnarray}

Now using variables~(\ref{y1y2}) together with Eqs.~(\ref{constraint}) and
(\ref{Hdot}) we can derive the corresponding evolution equations
for the variables $y_1$ and $y_2$ thus:
\begin{eqnarray}
\label{auto1}
& & \frac{{\rm d}y_1}{{\rm d}N}=
y_1 \left[ 3-3y_1+y_2+C(R) (1-y_1) \right]\,, \\
\label{auto2}
& & \frac{{\rm d}y_2}{{\rm d}N}=
y_2 \left[ -1-3y_1+y_2-C(R) y_1 \right]\,,
\end{eqnarray}
where $N \equiv {\rm ln}\,a$ and
\begin{eqnarray}
\label{CR}
C(R) \equiv \frac{R \dot{F}}{H(FR-f)}=
-3\frac{(FR-2f)F'R}{(FR-f)(F'R-F)}\,.
\end{eqnarray}
We also note that the following constraint equation also holds
\begin{eqnarray}\label{Ry1y2}
\frac{FR-2f}{FR-f}=-\frac{1-y_1-y_2}{2y_1} \,,
\end{eqnarray}
which shows that $R$ and thus $C(R)$ can in principle be 
expressed in terms of variables $y_1$ and $y_2$.

The behaviours of the variables $y_1$ and $y_2$ depend on the
behaviour of the function $C(R)$. 
In particular, the divergence of $C(R)$ can prevent $y_1$ 
and $y_2$ from reaching  equilibrium, as we shall see in the 
next Section. 
To proceed here we shall assume
that $C(R)$ is well behaved. In that case
an important step in understanding the dynamics of such systems
is to look at their equilibrium points/invariant sets and
their stabilities.
The fixed points $(y_1,y_2)$ satisfy
${\rm d}y_1/{\rm d}N=0={\rm d}y_2/{\rm d}N$.
In this case (even when $C(R)$ depends on $R$, 
but excluding the cases $C(R)=-3,-4$)
we obtain the following fixed points:
\begin{itemize}
\item $\PR:$ $(y_1, y_2)=(0, 1)$\,, $~~~~$ [We shall discuss this case further below.]

\item $\PM:$ $(y_1,y_2)=(0, 0)$\,,

\item $\PD:$ $(y_1, y_2)=(1, 0)$\,.

\end{itemize}

If $C(R)=-3$, as is the case with the model $f(R) \propto R^n$
($n \neq 1, 2$), we obtain (a) $(y_1, y_2)=(0, 1)$
and (b) $(y_1, y_2)=(y_1^{(c)}, 0)$ where $y_1^{(c)}$ is 
a constant.
When $C(R)=-4$, the fixed points are given by
(a) $(y_1, y_2)=(0, 0)$,
(b) $(y_1, y_2)=(1, 0)$ and (c)
$(y_1, y_2)=(y_1^{(c)}, 1-y_1^{(c)})$.
Note that in both these cases the latter points
in fact correspond to a line
of (rather than isolated) fixed points.

The stability of the fixed points in the 2-dimensional  
phase space $(y_1,y_2)$ can then be studied
by linearising the equations and obtaining the
eigenvalues of the corresponding Jacobian matrices calculated
at each equilibrium point \cite{CLW}. 
Assuming that ${\rm d}C/{\rm d}y_1$ and ${\rm d}C/{\rm d}y_2$ 
remain bounded, we find the following eigenvalues 
for the above fixed points:
\begin{itemize}
\item $\PR: ~(\lambda_1, \lambda_2)= (4+C(R), 1)$\,,

\item $\PM: ~(\lambda_1, \lambda_2)= (3+C(R), -1)$\,,

\item $\PD: ~(\lambda_1, \lambda_2)= (-3-C(R), -4-C(R))$\,.
\end{itemize}

In the following we shall also find it useful
to define an effective equation of state (EOS), $w_{\rm eff}$, which 
is related to the Hubble parameter via 
$\dot{H}/H^2=-(3/2)(1+w_{\rm eff})$.
Using Eq.~(\ref{Hdot}) we find 
\begin{eqnarray}
\label{weff}
w_{\rm eff}=-y_1+\frac13 y_2+\frac{\dot{F}}{3HF}
+\frac{\dot{\xi}}{3H\xi}-\frac{\dot{F}R}
{18F \xi H^3}\,.
\end{eqnarray}
Once the fixed points of the system are obtained, 
one can evaluate the corresponding effective 
equation of state by using this relation.

%-------------------------------------------------------------------
\section{Cosmological dynamics for models based on $f(R)$ theories}
\label{s3}
%-------------------------------------------------------------------
In this section we shall use the above formalism to study 
a number of families of 
$f(R)$ theories, recently considered in the literature. 

%-------------------------------------------------------------------
\subsection{$f(R)=R-\Lambda$}
%-------------------------------------------------------------------

The simplest model not ruled out by observations is the 
$\Lambda$CDM model. To start with, therefore, it is instructive 
to consider this model in this context, as it represents
the asymptotic state in a number of cases considered below.
The Lagrangian in this case is given by
\begin{eqnarray}
\label{act-lcdm}
f(R)=R-\Lambda\,,
\end{eqnarray}
which gives $C(R)=0$. 
Eqs.~(\ref{auto1}) and (\ref{auto2}) then reduce to
\begin{eqnarray}
\label{autoLam1}
& & \frac{{\rm d}y_1}{{\rm d}N}=
y_1 \left( 3-3y_1+y_2 \right)\,, \\
\label{autoLam2}
& & \frac{{\rm d}y_2}{{\rm d}N}=
y_2 \left( -1-3y_1+y_2 \right)\,, 
\end{eqnarray}
with fixed points given by
\begin{eqnarray}
\label{Lampoints}    
\PR: (y_1, y_2) =(0, 1)\,, \quad
\PM: (y_1, y_2) =(0, 0)\,, \quad
\PD: (y_1, y_2)=(1, 0)\,.
\end{eqnarray}
The corresponding eigenvalues can be evaluated to be
\begin{eqnarray}
\PR:~(\lambda_1, \lambda_2)=(1, 4)\,, \quad
\PM:~(\lambda_1, \lambda_2)=(3, -1)\,, \quad
\PD:~(\lambda_1, \lambda_2)=(-3, -4)\,.
\end{eqnarray}
Thus the fixed points $\PR$, $\PM$ and $\PD$ 
correspond to an unstable node, a saddle point 
and a stable node, respectively.
{}From Eq.~(\ref{weff}) the effective equation of state is
in this case given by $w_{\rm eff}=-y_1+y_2/3$.
This then leads to $w_{\rm eff}=1/3$, $0$ and $-1$
for the three points given in Eq.~(\ref{Lampoints}), 
indicating radiation-dominated,
matter-dominated and de-Sitter phases, respectively.
We have confirmed that this sequence of 
behaviours does indeed occur  
by directly solving the autonomous equations
(\ref{autoLam1}) and (\ref{autoLam2}).

The $\Lambda$CDM model corresponds to $f(R)=R-\Lambda$, 
in which case $C(R)$ vanishes because $F'=0$. 
Equation~(\ref{CR}) shows that $C(R)$ also vanishes when
\begin{eqnarray}
\label{con}
FR-2f=0\,,
\end{eqnarray}
provided that other terms in the expression of $C(R)$
do not exhibit divergent behaviour.
When $C(R) \to 0$ the system possesses the 
fixed points $P_r$, $P_m$ and $P_d$.
In principle, the nature of the point $P_d$ depends on the 
nature of the theory under study, i.e. the system 
(\ref{auto1})-(\ref{auto2}).
If the last terms in these equations vanish sufficiently fast as
$C(R) \to 0$, then $P_d$ corresponds to a
de-Sitter solution which is a stable node since its eigenvalues
are given by $(\lambda_1, \lambda_2)=(-3, -4)$.
If, on the other hand, these terms fall slower than
$1/N$, then they can contribute
to the evolution of $y_1$ and $y_2$ and the
point $P_d$ may be different from a de-Sitter point. 
We note, however, that for all the theories considered
in this paper $P_d$ corresponds to a de-Sitter point.

%-------------------------------------------------------------------
\subsection{$f(R)$ theories with the sum of power-law terms}
%-------------------------------------------------------------------

The families of models we shall consider in this section 
belong to the classes of theories  given by
\begin{eqnarray}
\label{both}
f(R)=R+\alpha R^{m}-\beta /R^n\,,
\end{eqnarray}
where $m$ and $n$ are real constants with the same sign and $\alpha$ and $\beta$
have dimensions [mass]$^{2(1-m)}$ and [mass]$^{2(n+1)}$ respectively.
Such theories have been 
considered with the hope of explaining both the early
and the late accelerating phases in the universe \cite{Meng04,Sot}.
In this subsection we shall make a detailed study of
models based on such theories in order
to determine whether they admit viable cosmological models,
with both early and late time acceleration phases.
In this case equation (\ref{Ry1y2}) becomes
\begin{eqnarray}
\label{gere}    
\frac{1-(m-2)\alpha R^{m-1}-(n+2)\beta R^{-n-1}}
{(m-1)\alpha R^{m-1}+(n+1)\beta R^{-n-1}}
=\frac{1-y_1-y_2}{2y_1}\,,
\end{eqnarray}
which in principle allows $R$ to be expressed
in terms of $y_1$ and $y_2$, at least implicitly.
The function $C(R)$ is given by 
\begin{eqnarray}
C(R)=-3 \frac{[1-(m-2)\alpha R^{m-1}-(n+2)\beta R^{-n-1}]
[m(m-1)\alpha R^m -n(n+1) \beta R^{-n}]}
{[1-m(m-2)\alpha R^{m-1}+n(n+2)\beta R^{-n-1}]
[(m-1)\alpha R^m+(n+1)\beta R^{-n}]}\,.
\end{eqnarray}

Before proceeding, some comments are in order concerning the 
variables $y_1$ and $y_2$. From the expression for Hubble function 
(\ref{Hubble}), one can observe that if $FR-f>0$ 
then one requires $6F\xi>0$,
as otherwise $H^2$ would be negative. 
Assuming (see below) that at late times
Eq.~(\ref{both}) tends to $R-\beta/R^n$, then $FR-f\rightarrow 
(1+n)\beta R^{-n}$ 
which is thus positive since observations 
(see section \ref{s31}) require $n>-1$ and $\beta>0$. 
Hence at late times $y_1$ and $y_2$ are positive and, 
from Eq.~(\ref{constraint}), their sum must be 
smaller than $1$, as otherwise $\rho_m$ would be negative. 
Similarly assuming (see below) that at early times Eq.~(\ref{both}) 
tends to $R+\alpha R^m$, then we have 
$FR-f\rightarrow (m-1)\alpha R^{m}$. 
To obtain an early time inflation, we need $m>1$ 
(as we shall see below).
Moreover, for the action to remain positive at early times, 
we require $\alpha>0$ 
which implies $f>0$. Therefore $FR-f$ is again positive. 
Thus $y_1$ and $y_2$ are positive also at early times,
with a sum which is smaller than $1$ 
for physical reasons. This indicates that at both early and late times, 
variables $y_1$ and $y_2$ can be treated as normalised.

%-------------------------------------------------------------------
\subsubsection{Limiting behaviours}
%-------------------------------------------------------------------
To study the full behaviour of the theories of the type
(\ref{both}) it is useful to first consider the limiting
cases where one of the nonlinear terms in the action will dominate.
We shall consider these cases separately.
\\

\noindent {\bf (i) The $\alpha=0$ case}
\\

In this case (\ref{both}) reduces to
\begin{eqnarray}
f(R)=R-\beta /R^n\,,
\end{eqnarray}
and $C(R)$ becomes
\begin{eqnarray}
 \label{CRal}
C(R) = 3n\frac{R^{1+n}-(2+n)\beta}
{R^{1+n}+n(2+n)\beta}\,,
\end{eqnarray}
which allows $R$ to be expressed in terms of 
$y_1$ and $y_2$ thus:
\begin{equation}
\label{R-eq}
R^{1+n}=\frac{\beta \left [3y_1+n(y_1-y_2+1)-y_2+1\right]}
{2y_1}\,.
\end{equation}
The critical points in this case are given by
\begin{itemize}
\item $P_r$:~$(y_1, y_2) =(0,1)$. 

Since the numerator and the denominator of Eq.~(\ref{R-eq}) 
tend to zero as one approaches $P_r$, care must be taken in this case.
We therefore split the analysis into three parts:
\begin{enumerate}
\item $P_{r1}$: This point corresponds to 
%$y_1\ll 1-y_2$, i.e. 
$\beta/R^{1+n} \ll 1$ 
and $C\rightarrow 3n$. 
The eigenvalues in this case
are given by $(\lambda_1,\lambda_2)=(3n+4,1)$ 
with $w_{\rm eff}=1/3$.

\item $P_{r2}$: This point corresponds to $\beta/R^{1+n} \gg 1$ 
and $C\rightarrow -3$. 
The eigenvalues in this case
are given by $(\lambda_1,\lambda_2)=(1,1)$ 
with $w_{\rm eff}=-2/3-1/n$.

\item $P_{r3}: R^{1+n} \to$ constant.
This case, however, does not occur since then
$y_1 \gg y_2-1$ (the reasoning 
would still remain the same if $y_1$ is of the same order
of magnitude as $1-y_2$) 
which would imply that $R^{n+1}$ should tend to 
$\beta (3+n)/2$. But from Eq.~(\ref{constraint}), one can deduce that 
$\kappa \rho_m\simeq (-FR+f)/2$ and from Eq.~(\ref{eq3}) 
that $\kappa \rho_m=-FR+2f$. 
Hence this would imply $(-FR+f)/2\simeq -FR+2f$ which is not possible.
\end{enumerate}
\item $P_m:~(y_1,y_2)=(0,0)$. 

In this case
$C \to 3n$, $(\lambda_1,\lambda_2)=(3(n+1),-1)$ and 
$w_{\rm eff}=0$.
\item $P_d:~(y_1,y_2)=(1,0)$.

In this case $R$ is a nonzero constant satisfying 
$R^{1+n}=(2+n)\beta$ with $C=0$, 
$(\lambda_1,\lambda_2)=(-3,-4)$ and $w_{\rm eff}=-1$.
This de-Sitter point exists for $n>-2$ provided that 
$R>0$ and $\beta>0$.
\item $P:~(y_1,y_2)=\left(-\frac{n+1}{n+3},0 \right)$.

In this case $C=3n$, $(\lambda_1,\lambda_2)=(1+1/n,-1)$ and
$w_{\rm eff}=-1-1/n$.

This fixed point is, however, not relevant in the 
asymptotic regimes that we shall consider.
This is because with observationally motivated parameters 
(see below) $\frac{n+1}{n+3} >0$ which thus implies 
$y_1<0$. However we have shown that for late and early 
times $0<y_1<1$.
\end{itemize}

A summary of fixed points in this case together with
their stability properties is given in Table \ref{tab1}.
An inspection of this table shows that 
the sequence of radiation ($P_{r1}$), matter ($P_{m}$) and 
de-Sitter acceleration ($P_{d}$) can be 
realized for $n>-1$.

%--------------------------------------------------
\begin{table}
\begin{center}
\begin{tabular}{|c||c|c|c|c|c|}
\hline
$n$ & $P_{r1}$ & $P_{r2}$ & $P_m$ & $P_d$ & $P$ \\
\hline
$n<-2$ & saddle & unstable & stable & -- & saddle \\
$-2< n<-4/3$ & saddle & unstable & stable & stable & saddle \\ 
$-4/3<n<-1$ & unstable & unstable & stable & stable & saddle \\
$-1<n<0$ & unstable & unstable & saddle & stable & stable \\
$n>0$ & unstable & unstable & saddle & stable & saddle \\
\hline
\end{tabular}
\caption{Fixed points and their natures for models 
based on theories of the type $f(R)=R-\beta /R^n$.
The existence of the $P_d$ point here assumes that
$\beta>0$. For negative $\beta$ the symbol `--' in the 
Table should be replaced by `stable' and vice versa.}
\label{tab1}
\end{center}
\end{table}
%--------------------------------------------------
\vskip 0.2in

\noindent {\bf (ii) The $\beta=0$ case}
\\

In this case (\ref{both}) reduces to
\begin{eqnarray}
f(R)=R+\alpha R^{m}\,.
\end{eqnarray}
The fixed points in this model can be obtained from the previous case
by simply taking $n\rightarrow -m$ and $\beta \rightarrow -\alpha$.
Hence, depending on the energy region the fixed points
can be summarised as follows.
\begin{enumerate}
\item High energy points ($\alpha R^{m-1} \gg 1$) \\
(1a) $P_{r2}$: $(y_1, y_2)=(0,1)$, $w_{\rm eff}=-2/3+1/m$
and $(\lambda_1,\lambda_2)=(1,1)$. \\
(1b) $P$: $(y_1, y_2)=\left(-\frac{1-m}{3-m},0\right)$, 
$w_{\rm eff}=-1+1/m$ and 
$(\lambda_1,\lambda_2)=(1-1/m,-1)$. 

\item Low energy points ($\alpha R^{m-1} \ll 1$) \\
(2a) $P_{r1}$: $(y_1, y_2)=(0,1)$, $w_{\rm eff}=1/3$
and $(\lambda_1,\lambda_2)=(4-3m,1)$. \\
(2b) $P_{m}$: $(y_1, y_2)=(0,0)$, 
$w_{\rm eff}=0$ and 
and $(\lambda_1,\lambda_2)=(3(1-m),-1)$. 

\item de-Sitter point ($\alpha R^{m-1}=1/(m-2)$) \\
(3) $P_d$: $(y_1, y_2)=(1,0)$, $w_{\rm eff}=-1$ and 
$(\lambda_1,\lambda_2)=(-3,-4)$.
\end{enumerate}
Note that the case $m=2$ requires a separate analysis since 
this corresponds to $C=-6$ independent of $R$ 
[see Eq.~(\ref{CRal})]. We shall return to this case below.
The fixed points and their properties for this case are summarised
in Table \ref{tab2}.\\
\\
%------------------------------------------------
\begin{table}
\begin{center}
\begin{tabular}{|c||c|c|c|c|c|}
\hline
$m$ & $P_{r1}$ & $P_{r2}$ & $P_m$ & $P_d$ & $P$ \\
\hline
$2<m$ & saddle & unstable & stable & stable & saddle \\
$4/3<m<2$ & saddle & unstable & stable & -- & saddle \\
$1<m<4/3$ & unstable & unstable & stable & -- & saddle \\
$0<m<1$ & unstable & unstable & saddle & -- & stable \\
$m<0$ & unstable & unstable & saddle & -- & saddle \\
\hline
\end{tabular}
\caption{Fixed points and their natures for
models based on theories of the type
$f(R)=R+\alpha R^{m}$.
The existence of the $P_d$ point here assumes that 
$\alpha>0$. For negative $\alpha$ the symbol `--' in the   
Table should be replaced by `stable' and vice versa.}
\label{tab2}
\end{center}
\end{table}
%------------------------------------------------------
%
One can in principle consider the fixed points in
Tables \ref{tab1} and \ref{tab2} as respectively being relevant at
the future and the past respectively.
Using the above asymptotic
information we shall now consider
families of theories of this type separately.

%-------------------------------------------------------
\subsubsection{Theories of type $f(R)=R-\beta / R^{n}$}
%-------------------------------------------------------

In the metric approach, models based on
theories of this type were considered
in Refs.~\cite{Capo03,Carroll03} as ways of
producing a late-time acceleration of the universe.
In addition to the difficulties such theories face
(as discussed in the Introduction),
they have recently been shown to be unable to
produce a standard matter-dominated era followed by
an accelerated expanding phase for $n>0$ \cite{APT}.
Using the Palatini approach, however, such theories can be shown
to be able to give rise to such a standard matter-dominated
phase, as we shall see below.

\begin{figure}
\includegraphics[height=3.0in,width=3.0in]{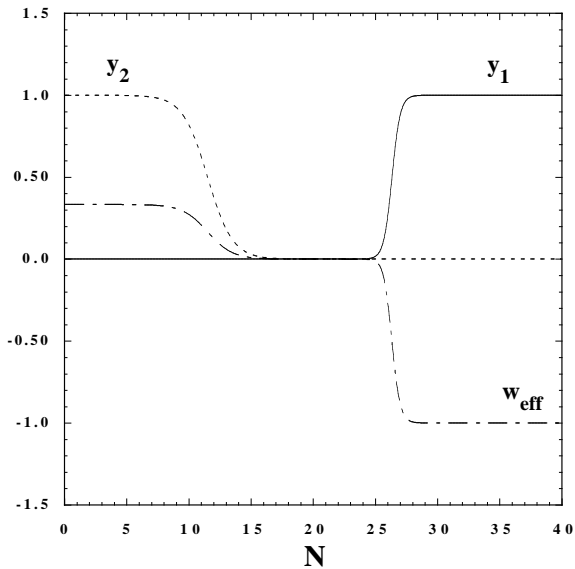}
\caption{
Evolutions of the variables $y_1$ and $y_2$
for the model $f(R)=R-\beta/R^n$ with $n=0.02$, 
together with the effective 
equation of state $w_{\rm eff}$.
Initial conditions were chosen to be
$y_1=10^{-40}$ and $y_{2}=1-10^{-5}$.
This demonstrates that the sequence of
($P_{r1}$) radiation-dominated ($w_{\rm eff}=1/3$),
($P_{m}$)  matter-dominated ($w_{\rm eff}=0$)
and ($P_{d}$) de-Sitter acceleration ($w_{\rm eff}=-1$) eras
occur for these models.}
\label{fig1}
\end{figure}

To proceed we shall, for the sake of compatibility
with the observational constraints obtained
in the following sections, confine ourselves
to the cases with $n>-1$ and $\beta>0$. In such cases,
Table \ref{tab1} shows that $P_{r1}$ is an unstable node (source), 
$P_m$ is a saddle and $P_d$
is a stable node. Numerical simulations confirm
that such theories indeed admit the 3 post-inflationary phases,
namely radiation-dominated,
matter-dominated and de-Sitter phases (see Fig.~\ref{fig1}).
The point $P$ can never be reached  because, for the
considered range of $n$, one can show that the positive
variable $y_1$ is an increasing function of time, i.e. 
${\rm d}y_1/{\rm d}N>0$ \footnote{To show that ${\rm d}y_1/{\rm d}N>0$, 
we solve ${\rm d}y_1/{\rm d}N=0$ with respect to $n$ to obtain $n(y_1,y_2)$. 
Then plotting the function $n(y_1,y_2)$ in the physical range of 
variables $(y_1,y_2) \in\left[0,1\right]$, 
we find that the set of triplets 
$(n,y_1,y_2)$, such that ${\rm d}y_1/{\rm d}N=0$, 
all satisfy $n<-1$ apart from the special 
(and therefore the zero measure) set of
fixed points which are also solutions of ${\rm d}y_1/{\rm d}N=0$ 
for any $n$.}.
The point $P$, however,  occurs for $y_1=0$ and hence can never be 
reached except by taking $y_1<0$ which is unphysical since it implies
a negative $H^2$.

%%%%%%%%%%%%%%%%%%%%%%%%%%%%
%%%%%%%%%%%%%%%%%%%%%%%%%%%%
 
The initial value of the ratio $r=y_1(0)/y_2(0)$ plays
an important role in determining the duration of the matter-dominated phase.
This ratio is related to cosmological parameters such as the 
present value of $\Omega_{m}$
that will be constrained by observations in the next section.
For the existence of a prolonged matter-dominated epoch we require the 
condition $r\ll 1$.
In this case the smaller the value of $r$ the longer will be
the matter-dominated phase.
For large enough values of $r$, the solutions directly 
approaches the stable de-Sitter point $P_{d}$
after the end of the radiation-dominated era.

%------------------------------------------------------------------------------
\subsubsection{Theories of type $f(R)=R+\alpha R^{m}-\beta /R^n$}
\label{subgenemodel}
%------------------------------------------------------------------------------

In this section, we shall employ the above asymptotic analyses
to study the dynamics of the generalised theories of this type.
In particular we shall give a detailed argument which indicates
that, under some general assumptions, an initial inflationary 
phase cannot be followed by a radiation phase.

We shall proceed by looking at the early and late dynamics
separately. In this connection it is important to decide which
phases can be considered to be produced by the early and late 
time approximations
$f(R)=R+\alpha R^{m}$ and $f(R)=R-\beta/R^n$ respectively.
We note that in the discussion below we shall at times 
divide the `early' phase into high energy 
($\alpha R^{m-1} \gg 1$) and low energy 
($\alpha R^{m-1} \ll 1$) regimes.
To proceed we start by 
assuming that the early time approximation covers inflation and
radiation-dominated phases and the late time approximation covers 
the matter-dominated and
dark energy phases. Thus our analysis does not
cover the epochs where the two nonlinear terms are 
comparable. Such epochs can occur between early and late
times or at late times. Nevertheless our assumption regarding early
times should still be valid since $R$ should be large in such regimes.
Also observations still require that $n>-1$ and $\beta>0$.

Table \ref{tab2} shows that to have an inflationary stage
followed by a radiation-dominated phase 
we must have $m>4/3$, since this is the only way to obtain 
a saddle radiation phase $P_{r1}$ 
(which would be required to obtain an exit from this phase).
The de-Sitter point $P_{d}$ exists for $m>2$, 
but one can not use this fixed point for inflation 
followed by the radiation era because $P_d$ is a stable attractor
which does not admit the exit from inflation.
The inflationary epoch can only come from the point 
$P_{r2}$ with $m>3$ or point $P$ with $m>3/2$, 
in order to ensure that $w_{\rm eff}<-1/3$.

Let us consider the case $m>2$. Taking into account the fact that 
$P_{r2}$ is in a high energy region ($\alpha R^{m-1} \gg 1$), 
the possible trajectory in this case is $P_{r2} \to P \to P_d$,
since $P_{r1}$ and $P_{m}$ are in the low energy region.
All points $P_{r2}$, $P$ and $P_d$ correspond
to inflationary solutions for $m>3$, whereas $P_{r2}$
is not an inflationary solution for $2<m<3$.
In neither case do we have a successful exit from 
the de-Sitter epoch $P_{d}$ to the radiation phase.

\begin{figure}
\includegraphics[height=3.0in,width=3.0in]{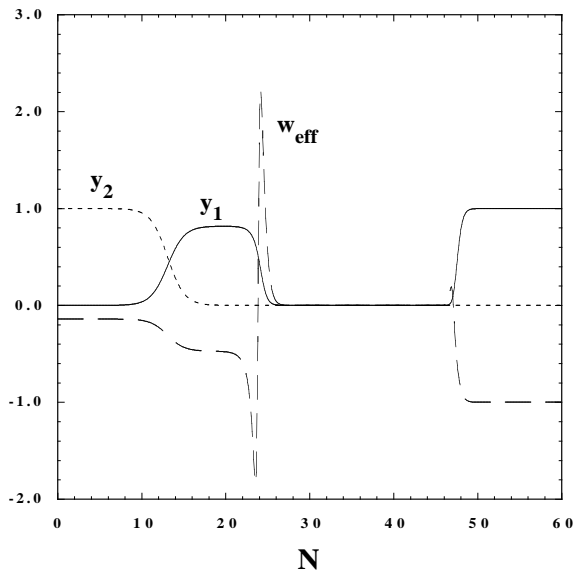}
\caption{Evolution of $y_1$ and $y_2$ together with
the effective equation of state $w_{\rm eff}$
for the model $f(R)=R+\alpha R^m -\beta/R^n$
with $m=1.9$, $n=1$, $\alpha=1$, $\beta=10^{-60}$. 
The initial conditions are 
$y_{1}=1.71 \times 10^{-6}$ and $y_{2}=1-2.09 \times 10^{-6}$.
The system starts from a non-standard radiation-type phase
(corresponding to the $P_{r2}$ point with $w_{\rm eff}=-2/3+1/m=-0.14$) 
followed by an inflationary phase (corresponding to 
the $P$ point with $w_{\rm eff}=-1+1/m=-0.47$)
and then followed by a matter-dominated phase
(corresponding to the $P_m$ point with $w_{\rm eff}=0$).
The system finally approaches a de-Sitter point $P_{d}$
because of the presence of the $\beta/R^n$ term.
}
\label{fullmodel}
\end{figure}

When $4/3<m<2$ it is possible to have the sequence: 
(i) $P_{r2} \to P \to P_m$, or
(ii) $P_{r2} \to P_{r1} \to P_{m}$.
Note that $P_{r2}$ corresponds to a non-standard evolution 
with $-1/6<w_{\rm eff}<1/12$.
When $m>3/2$ the point $P$ leads to an accelerated expansion, 
but the case (i) is not viable because of the absence of the 
radiation era after inflation. Similarly, the case (ii) 
can be ruled out since it does not possess an
inflationary phase. Now since both $P$ and $P_{r1}$ 
are saddle points, 
we may ask whether the sequence $P_{r2} \to P \to P_{r1} \to P_m$
can occur. To study this possibility, we proceeded numerically
and solved the autonomous equations for a range of initial conditions
close to $y_{1}=0$ and $y_{2}=1$ and a range of model parameters. 
Despite extensive numerical searches
we were unable to find a radiation-dominated phase  
between the de-Sitter and matter epochs. 
Fig.~\ref{fullmodel} gives a typical example of such simulations
showing the evolution of the variables
$y_1$ and $y_2$ together with $w_{\rm eff}$.
Clearly the radiation epoch is absent between the de-Sitter and 
matter epochs. The sequence (i) $P_{r2} \to P \to P_m$ is, 
however, realized 
as expected, which is finally followed by a de-Sitter universe
as the $\beta/R^n$ term becomes important.
Thus the models with $3/2<m<2$ seem to be unable to produce
an initial inflationary phase leading to a radiation-dominated
epoch. 
\\
The above discussion indicates that the assumption $m>3/2$, which
is necessary for an early-time acceleration phase
to occur, is incompatible with the subsequent
radiation-dominated phase.
\\

To complete this analysis we also consider the range
where $1<m<4/3$. In this case there are two possible sequences:
(i) $P_{r2} \to P \to P_{m}$ and 
(ii) $P_{r1} \to P_{m}$.
In the case (i) the system starts from a non-standard
high energy phase $P_{r2}$ ($\alpha R^{m-1} \gg 1$), 
which is followed by 
a non-inflationary phase $P$. Since neither an inflationary phase nor
a radiation-dominated epoch exist, the sequence (i) is therefore not 
cosmologically viable.
In the case (ii) the system starts from a low-energy 
radiation-dominated phase ($\alpha R^{m-1} \ll 1$) and is 
followed by a matter-dominated epoch.
As long as the term $\beta/R^n$ becomes important at late
times, the sequence (ii) is followed by a de-Sitter point.
Thus the case (ii) can at least account for the last three phases
required in cosmology.
When $0<m<1$ it is also possible to have the sequence 
$P_{r1} \to P_{m} \to P_{d}$.

%%%%%%%%%%%%%%%%%%%%%%%%%%%%
%%%%%%%%%%%%%%%%%%%%%%%%%%%%

Finally, we shall consider the case where the
term $\alpha R^m$ is comparable to $R$ around the present epoch.
This case is suggested by the likelihood analysis considerations of the 
model parameters in the next section. 
When $0<m<1$ this case is similar to the one for the model 
$f(R)=R-\beta/R^n$ with $-1<n<0$. 
If $m>1$ this corresponds to a non-standard cosmological evolution
with $\alpha R^m \gg R$ in the past, which can be realized for 
a very large coupling $\alpha$.
To complement the discussion above,
we shall also briefly consider this case.
To start with let us consider the case $m \ne 2$.
Since $\alpha R^{m-1} \gg 1$ prior to the dark energy epoch, the 
fixed points correspond to either
(i) $P_{r2}$ with an EOS $w_{\rm eff}=-2/3+1/m$, 
eigenvalues $(\lambda_1, \lambda_2)=(1,1)$ or 
(ii) $P$ with an EOS $w_{\rm eff}=-1+1/m$, 
eigenvalues $(\lambda_1, \lambda_2)=(1-1/m,-1)$. 
Since we are considering the case $m>1$, we have 
(i) $w_{\rm eff}<1/3$
and (ii) $w_{\rm eff}<0$.
This shows that the radiation-dominated phase is not realized 
unless $m$ is unity (i.e. Einstein gravity)

We shall consider the case $m=2$ in the next subsection.

%-------------------------------------------------------
\subsubsection{The special case $m=2$}
\label{m2sin}
%-------------------------------------------------------

\noindent To conclude our consideration of theories of 
the type (\ref{both}), we shall consider the special case of
these theories with $m=2$, which have
recently attracted some attention in the literature \cite{MenWan04}.
\\

\noindent{\bf (A) Theories of the type $f(R) = R+\alpha R^2$}\\
\\
In this case we have $C=-6$.
The critical points, eigenvalues and the effective EOS are

\begin{itemize}
\item $P_r:~ (y_1, y_2) =(0,1)$, $(\lambda_1,\lambda_2)=(-2,1)$. 

We note that in this case $w_{\rm eff}$ is undefined, 
as can be seen from Eq.~(\ref{weffape2}) in the Appendix.
This, however, does not play an important role in the 
reasoning below, as we shall see.

\item $P_m=(0,0)$, $(\lambda_1,\lambda_2)=(-3,-1)$, 
$w_{\rm eff}=0$.

This matter point exists only in the region $\alpha R \ll 1$
because of the relation $\alpha R=2y_1/(1-y_1-y_2)$.

\item $P_d=(1,0)$, $(\lambda_1,\lambda_2)=(2,3)$, $w_{\rm eff}=0$.

This point exists only in the region $\alpha R \gg 1$.
\end{itemize}

An important feature of this model is that the fixed point $P_d$ 
in this case is an unstable node mimicking a dust Universe. 
Depending upon whether the radiation ($y_1 \ll y_2$) 
or the dark energy ($y_2 \ll y_1$) dominates initially, 
the universe starts with the EOS 
of matter ($P_d$) or radiation ($P_r$) and ends 
in a matter era ($P_m$) which is a stable node. 
Before reaching the final  attractor $P_m$, there can 
exist a time $N=N_s$ 
such that $y_1=(1-y_2)/3$. In particular this always occurs
when $y_1>1/3$ at early times, as would be expected 
in a universe dominated by dark energy.
In this case, the EOS diverges at $N_s$ as we see 
from Eq.~(\ref{weffape2}) in the Appendix.
Note that when $y_1=(1-y_2)/3$, the scalar curvature is regular 
($\alpha R=1$) with finite $y_1$ and $y_2$,
indicating that there is no physical singularity, despite the singular 
behaviour of the effective EOS.
This can be understood by recalling that the $w_{\rm eff}$ 
so defined is just a mathematical construction 
for $f(R)$ theories.
\\

\noindent{\bf (B) Theories of the type $f(R) = R+\alpha 
R^2-\beta/R$}\\
\\
Let us consider the case in which the terms $\alpha R^2$ and $\beta/R$ are
important for early and late times, respectively.
Then as long as $\alpha>0$ and $\beta>0$, it is possible to have the sequence 
$P_d \to P_{r} \to P_m$ followed by a de-Sitter point $P_{d}$
which appears because of the presence of the term $\beta/R$.
However since $P_r$ corresponds to a divergent EOS, we do not have 
a standard radiation-dominated epoch in this model.
\\
Let us next consider the case in which the signs of $\alpha$ and 
$\beta$ are different, say $\alpha>0$ and $\beta<0$.
We will encounter this situation when we carry out a 
likelihood analysis of model parameters in the next section.
Then it can happens that $C(R)$ diverges as $R$ approaches either
$(-2\beta/\alpha)^{1/3}$ or $R \to \sqrt{-3\beta}$.
In this case the effective EOS diverges and 
${\rm d}y_1/{\rm d}N$ approaches $-\infty$.
For the model parameters constrained by observations
the divergence occurs at $R=(-2\beta/\alpha)^{1/3}$ 
before reaching $R=\sqrt{-3\beta}$, since in that case 
$(-2\beta/\alpha)^{1/3}>\sqrt{-3\beta}$.
We also note that the radiation-dominated phase is absent in this case 
as well if we go back to the past.

This provides an example of what can happen when $C(R)$ 
diverges, which may also occur for other values of $(m,n)$.

%-------------------------------------------------------
\subsection{Theories of type $f(R)=R+\alpha {\rm ln}\,R-\beta$}
%-------------------------------------------------------

Finally we shall consider theories of the type
\begin{equation}
f(R)=R+\alpha {\rm ln}\,R-\beta\,.
\end{equation}
The cosmological evolution of the models based
on these theories has
been studied~\cite{NojOdi04,MenWan04A} and it
has been claimed that such theories 
have a well-defined Newtonian limit \cite{NojOdi04}.
{}For these theories Eqs.~(\ref{CR}) and (\ref{Ry1y2}) give
\begin{eqnarray}
\label{C-log}
C(R)=3\alpha \frac{R-\alpha+2\alpha {\rm ln}\,R-2\beta}
{(\alpha-\alpha {\rm ln}\,R +\beta)(R+2\alpha)}\,,
\end{eqnarray}
and
\begin{eqnarray}
\frac{R-\alpha +2\alpha {\rm ln}\,R -2\beta}
{\alpha -\alpha {\rm ln}\,R +\beta}=
\frac{1-y_1-y_2}{2y_1}\,.
\label{conlog}
\end{eqnarray}
These imply $C \to 0$ in the limit $R \to \infty$
and $C \to -3$ in the limit $R \to 0$.
We also note that assuming the denominator in (\ref{C-log})
is non-zero, then $C=0$ for a constant $R$ satisfying 
$R-\alpha+2\alpha {\rm ln}\,R-2\beta=0$.
Using the constraint equation (\ref{conlog}), the 
critical points can be found to be
\begin{enumerate}
\item $P_{r}:~(y_1,y_2)=(0,1)$.

\begin{enumerate}
\item $P_{r1}$: $R \to \infty$ and $C\to 0$. 
The eigenvalues are given by $(\lambda_1,\lambda_2)=(4,1)$ 
with $w_{\rm eff}=1/3$.

\item $P_{r2}$: $R \to 0$ with $C\to -3$.
The eigenvalues are given by $(\lambda_1,\lambda_2)=(1,1)$ 
with $w_{\rm eff} \to -\infty$.
\end{enumerate}
\item $P_m:~(y_1,y_2)=(0,0)$. 

$R \to \infty$, $C \to 0$, $(\lambda_1,\lambda_2)=(3,-1)$ and 
$w_{\rm eff}=0$.
\item $P_d:~(y_1,y_2)=(1,0)$.

$C=0$, $(\lambda_1,\lambda_2)=(-3,-4)$ and 
$w_{\rm eff}=-1$, with $R$
behaving as a constant near the critical point.
\item $P:~(y_1,y_2)=(-1/3, 0)$.

$R \to 0$, $C=-3$, $(\lambda_1,\lambda_2)=(0,-1)$ and 
$w_{\rm eff} \to -\infty$.

\end{enumerate}

{}From the above discussion
it is clear that one can obtain the sequence of 
radiation-dominated ($P_{r1}$, 
matter-dominated ($P_{m}$), and 
de-Sitter ($P_{d}$) eras.
We have checked numerically that this sequence
indeed occurs.
Note that the points $P_{r2}$ and $P$ are irrelevant to 
realistic cosmology, since the system approaches the stable 
de-Sitter point $P_{d}$ with a constant $R$ before 
reaching $R \to 0$.

%-----------------------------------------------------------------------------%
\section{Confronting $f(R)$ gravity models with observational data} 
\label{s4}
%-----------------------------------------------------------------------------%

The detailed phase space analysis given in the previous Section
provides a clear picture of the possible dynamical modes of behaviour 
that can occur for models based on these theories. 
These analyses demonstrate 
that none of the theories considered here can produce all four phases
required for cosmological evolution. They, however, show that a subset
of these theories are capable of producing the last three phases,
i.e. radiation-dominated, matter-dominated and late time accelerating
phases. As was mentioned above the occurrence of these phases 
is a necessary but not a sufficient
condition for cosmological viability of such theories.
To be cosmologically viable, it is also necessary
that the subset of parameters in these theories that allow these
phases to exist are also compatible with
observations. 

In this Section we shall study this observational compatibility 
by confronting models based on these theories with observational data.
To proceed, we rewrite equations (\ref{eq3}), 
(\ref{baeq3}) and (\ref{Hubble}), using the redshift
parameter $z=a_0/a-1$ and employing the 
expressions $\rho_m=\rho_{m0}(1+z)^3$ and
$\rho_r=\rho_{r0}(1+z)^4$ to obtain
\begin{eqnarray}
\label{zeq1}	
& & FR-2f=-3H_0^2 \Omega_{m0} (1+z)^3\,, \\
\label{zeq2}
& & \frac{{\rm d}R}{{\rm d}z}=
-\frac{9H_0^2 \Omega_{m0}(1+z)^2}{F'R-F}\,, \\
\label{zeq3}
& & \frac{H^2}{H_0^2}=
\frac{3\Omega_{m0}(1+z)^3+6\Omega_{r0}(1+z)^4+f/H_0^2}
{6F\left[ 1+\frac{9H_{0}^2F' \Omega_{m0}(1+z)^3}
{2F(F'R-F)} \right]^2}\,,
\end{eqnarray}
where $\Omega_{m0} \equiv \kappa \rho_{m0}/3H_0^2$
and $\Omega_{r0} \equiv \kappa \rho_{r0}/3H_0^2$ and 
a subscript $0$ denotes evaluation at the present time.
{}From Eq.~(\ref{zeq1}) or (\ref{zeq2}) one can express
$R$ in terms of $z$ for a given $f(R)$ model.
Equation~(\ref{zeq3}) can then be used to obtain the Hubble parameter
in terms of $z$.

We note that for a given $f(R)$ theory sourced by cold dark matter,
$\Omega_{m0}$ is uniquely determined once 
units are chosen such that $H_0=1$ \cite{AmaElg06}.
If radiation is also present, then
both $H_0$ and $\Omega_{r0}$ need to be 
specified in order to determine $\Omega_{m0}$ uniquely.
In the following we choose $\Omega_{r0}=5\times 10^{-5}$ and $H_0=1$
(see below).

To constrain $f(R)$ models, we shall use three sets of data:
\begin{itemize}
\item (I) The first year data set
from the Supernova Legacy Survey (SNLS) \cite{Ast06} 
with $71$ new supernovae below $z=1.01$, together with another $44$ 
low $z$ supernovae already available, i.e. a total of 115 SNe.

\item (II) The Baryon acoustic oscillation peak (BAO) 
recently detected in the correlation function of luminous red 
galaxies (LRG) in the Sloan Digital Sky Survey \cite{Eis05}.
This peak corresponds to the first acoustic peak at recombination
and is determined by the sound horizon. The observed scale of
the peak effectively constrains the quantity
\begin{equation}
\label{BAOcon}
A_{0.35}=D_V(0.35)\frac{\sqrt{\Omega_{m0}H_0^2}}{0.35}
=0.469\pm 0.017\,,
\end{equation}
where $z=0.35$ is the typical LRG redshift
and $D_V(z)$ is the comoving angular diameter distance defined as
\begin{equation}
D_V(z)=\left[D_M(z)^2\frac{z}{H(z)}\right]^{1/3}\,,\quad
{\rm with} \quad 
D_M=\int^z_0\frac{{\rm d}z}{H}\,.
\end{equation}

\item (III) The CMB shift parameter. 
This constraint is defined by
\begin{equation}
\label{CMBcon}
R_{1089}=\sqrt{\Omega_{m0}H_0^2}\int^{1089}_0 
\frac{{\rm d}z}{H}=1.716\pm 0.062\,,
\end{equation}
which is a measure of the distance between $z=0$ and $z=1089$. 
It relates the angular diameter distance to the last scattering 
surface with the angular scale of the first acoustic peak.
An important feature of this parameter is 
that it is model independent and insensitive to 
perturbations \cite{Maar-etal}.
To be able to use it one has to have a 
standard matter-dominated era at decoupling.
\end{itemize}

For the goodness of fit we employ the standard 
$\chi^2$ minimization, defined by
$$
\chi^2=\sum_{i=1}^{n}\frac{(m^{{\rm obs}}_i-m_i^{{\rm th}})^2}{\sigma^2_i}\,,
$$
where $n$ is the number of data points and $m_i^{{\rm obs}}$
and $m_i^{{\rm th}}$ are respectively
the observed and the theoretical magnitudes calculated from the model. 
We shall marginalise the Hubble constant by defining as usual 
a new $\chi^2$:
$$
\bar \chi^2=-2\ln\int^{+\infty}_{-\infty}e^{-\chi^2/2}
{\rm d}\bar M\,,
$$
where $\bar M=M-5\ln (H_0)+25$ and 
$M$ is the magnitude zero point offset. 
After some algebraic manipulations and defining 
\begin{eqnarray}
A=\sum_{p=1}^{115}\frac{(m^{{\rm obs}}_i-5 \ln (D_l))^2}{\sigma^2_i}\,,
\quad
B=\sum_{p=1}^{115}\frac{m^{{\rm obs}}_i-5 \ln (D_l)}{\sigma^2_i}\,,
\quad
C=\sum_{p=1}^{115}\frac{1}{\sigma^2_i}\,,
\end{eqnarray}
we obtain 
\begin{eqnarray}
\bar \chi^2=A-\frac{B^2}{C}+\ln \frac{C}{2\pi}\,.
\end{eqnarray}
Finally we note that the BAO and the CMB shift parameter do not depend on 
the Hubble constant. 
Having only one measure point in each case, these tests can be used 
in the same way as priors.

In what follows we shall proceed to place observational constraints
on the models discussed in the previous section.
Our choice of units such that $H_0=1$ is 
consistent with marginalisation over $H_0$,
similar to the choice made in the literature \cite{AmaElg06}.

%----------------------------------------------------
\subsection{Theories of the type $f(R)=R-\beta /R^n$ } \label{s31}
%------------------------------------------------------

We shall first consider theories of the type
\begin{eqnarray}
\label{model1}	
f(R) = R-\beta/R^n\,.
\end{eqnarray}
As we already demonstrated in the previous section, one can obtain
in this case a sequence of radiation-dominated, matter-dominated
and acceleration phases provided 
that $n>-1$.

\begin{figure}
\includegraphics[width=8cm]{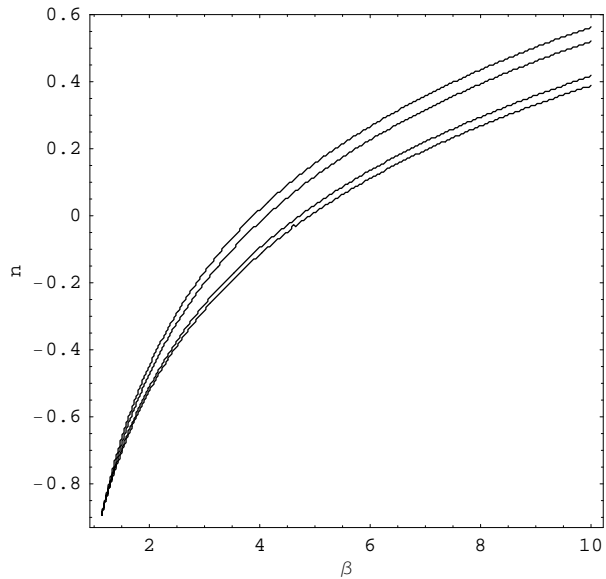}
\caption{ The $68.3\%$ and $95.4\%$ 
confidence levels for the model
based on the theory $f(R)=R-\beta/R^n$, 
constrained by SNLS data only.} 
\label{snonly}
\end{figure}

Recently Amarzguioui {\it et al.} \cite{AmaElg06}
have found that models of this kind are 
compatible with the supernova `Gold' data set
from Riess {\it et al.} \cite{Rie04} together with the
BAO constraint (\ref{BAOcon}) and the CMB
constraint (\ref{CMBcon}), subject to constraints
on the values of the parameters $\beta$ and $n$.
In their combined analysis the best-fit model was found 
to be $\beta=3.6$ and $n=-0.09$.
Here we extend this work by employing the more 
recent first year Super-Nova Legacy Survey (SNLS) data 
as well as taking into account the presence of radiation.
In Fig.~\ref{snonly} we show observational contour plots
at the $68\%$ and $95\%$ confidence levels
obtained from the SNLS data only.

%%%%%%%%%%%%%%%%%%%%%%%%%%%%
%%%%%%%%%%%%%%%%%%%%%%%%%%%%

In this case the best-fit value (smallest $\chi^2$) 
is found to be $\chi^2=116.55$ with 
$\beta=12.5$ and $n=0.6$. This is consistent with
the results obtained in Ref.~\cite{AmaElg06}, namely $\beta=10.0$
and $n=0.51$ and with the results obtained in the previous section
according to which the transition from the 
matter-dominated era to the de-Sitter era occurs for $n>-1$.
As $n$ gets closer to $-1$, it is difficult to realize 
a late-time acceleration phase since the model (\ref{model1})
approaches Einstein gravity.

\begin{figure}
\includegraphics[width=8cm]{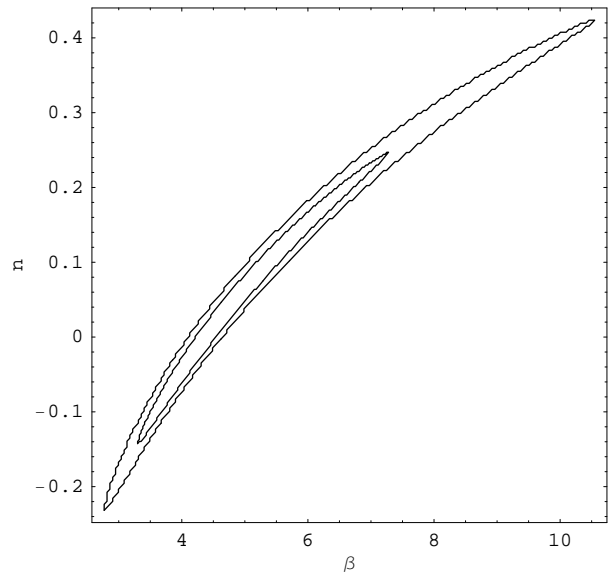}
\caption{The $68.3\%$ and $95.4\%$ confidence levels for the model
based on the theory $f(R)=R-\beta /R^n$,
constrained by SNLS, BAO and CMB data.} 
\label{forR-Rn}
\end{figure}

If we further include constraints from BAO and CMB, 
the likelihood parameter space is reduced significantly 
as seen from Fig.~\ref{forR-Rn},
with narrower ranges of the allowed values of
$n$ and $\beta$ lying in the intervals $n\in\left[-0.23,0.42\right]$ and 
$\beta\in\left[2.73,10.6\right]$ at the $95\%$ confidence level.
This difference is mainly due to the fact that the BAO data better 
constrains the cold dark matter parameter than the SN data alone. 
The best-fit values are found to be $\beta=4.63$ and 
$n=0.027$ with $\chi^2=116.69$.

%%%%%%%%%%%%%%%%%%%%%%%%%%%%
%%%%%%%%%%%%%%%%%%%%%%%%%%%%

%%%%%%%%%%%%%%%%%%%%%%%%%%%%
\begin{figure}[h]
\centering
\includegraphics[width=12cm]{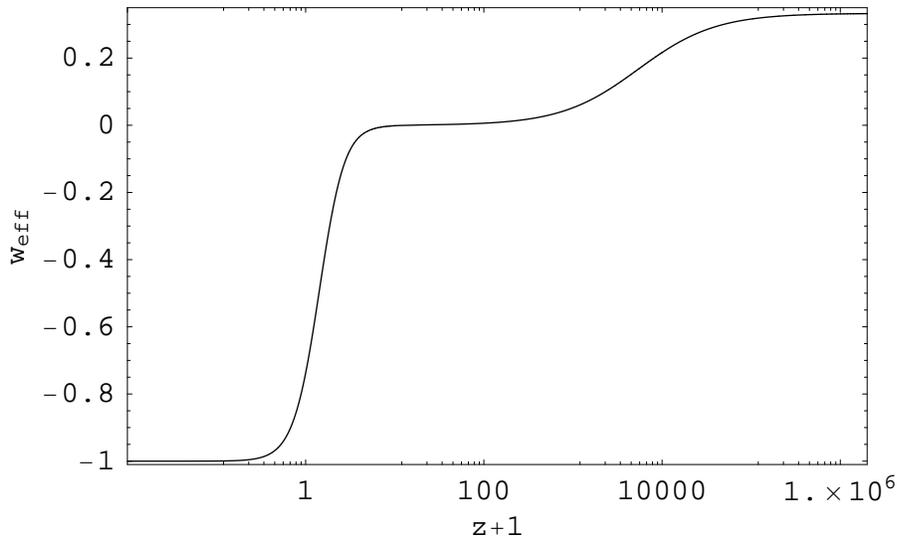}
\caption{\label{forR-RnDif} 
Evolution of the effective equation of state $w_{\rm eff}$,
for the model based on the theory $f(R)=R-\beta /R^n$,
with the best-fit value $\beta=4.63$ and $n=0.027$.
For very large $z$, $w_{\rm eff}=1/3$, 
which then decreases and approaches $0$ at redshift of few tens.
The accelerated expansion occurs around $z=1$ after which
$w_{\rm eff}$ becomes smaller than $-1/3$.
In the future $w_{\rm eff}$ asymptotically 
approaches $-1$. }
\end{figure}
%%%%%%%%%%%%%%%%%%%%%%%%%%%%

Note that the $\Lambda$CDM model corresponding to $\beta=4.38$
and $n=0$ lies in the $68\%$ confidence level.
In Fig.~\ref{forR-RnDif}, we have plotted the evolution of $w_{\rm eff}$
for the best-fit case.
This shows that the models in agreement with the data undergo a transition 
from a radiation-dominated to a matter-dominated epoch followed by 
an acceleration phase which in the future
asymptotically approaches a de-Sitter solution. 
This is a nice realisation of the trajectories that follow the sequence 
$P_{r1}\to P_{m} \to P_{d}$ described 
in the previous section. 

%---------------------------------------------------------
\subsection{Theories of the type $f(R)=R+\alpha R^m-\beta/R^n$} 
\label{s32}
%--------------------------------------------------------

We next consider theories of the type 
\begin{eqnarray}
\label{fRobser}
f(R) = R+\alpha R^m -\beta/R^n\,.
\end{eqnarray}
In the previous section we showed that an initial inflationary
phase ( requiring $m>3/2$) 
does not seem to be compatible with a 
subsequent radiation-dominated epoch.
However, dropping the requirement of an inflationary phase,
such theories can produce the last 3 phases, namely 
radiation-dominated, matter-dominated and 
late acceleration epochs,
provided the radiation-dominated era 
starts in the $\alpha R^{m-1} \ll 1$ regime.
If the term $\alpha R^m$ becomes smaller than $\beta/R^n$ prior to 
the decoupling epoch, observational constraints on the
models based on
(\ref{fRobser}) are similar to those on the
models based on (\ref{model1}),
i.e., $n \in\left[-0.23,0.42\right]$ and $\beta \in\left[2.73,10.6\right]$ 
at the $2\sigma$ level.

\begin{figure}
\includegraphics[width=8cm]{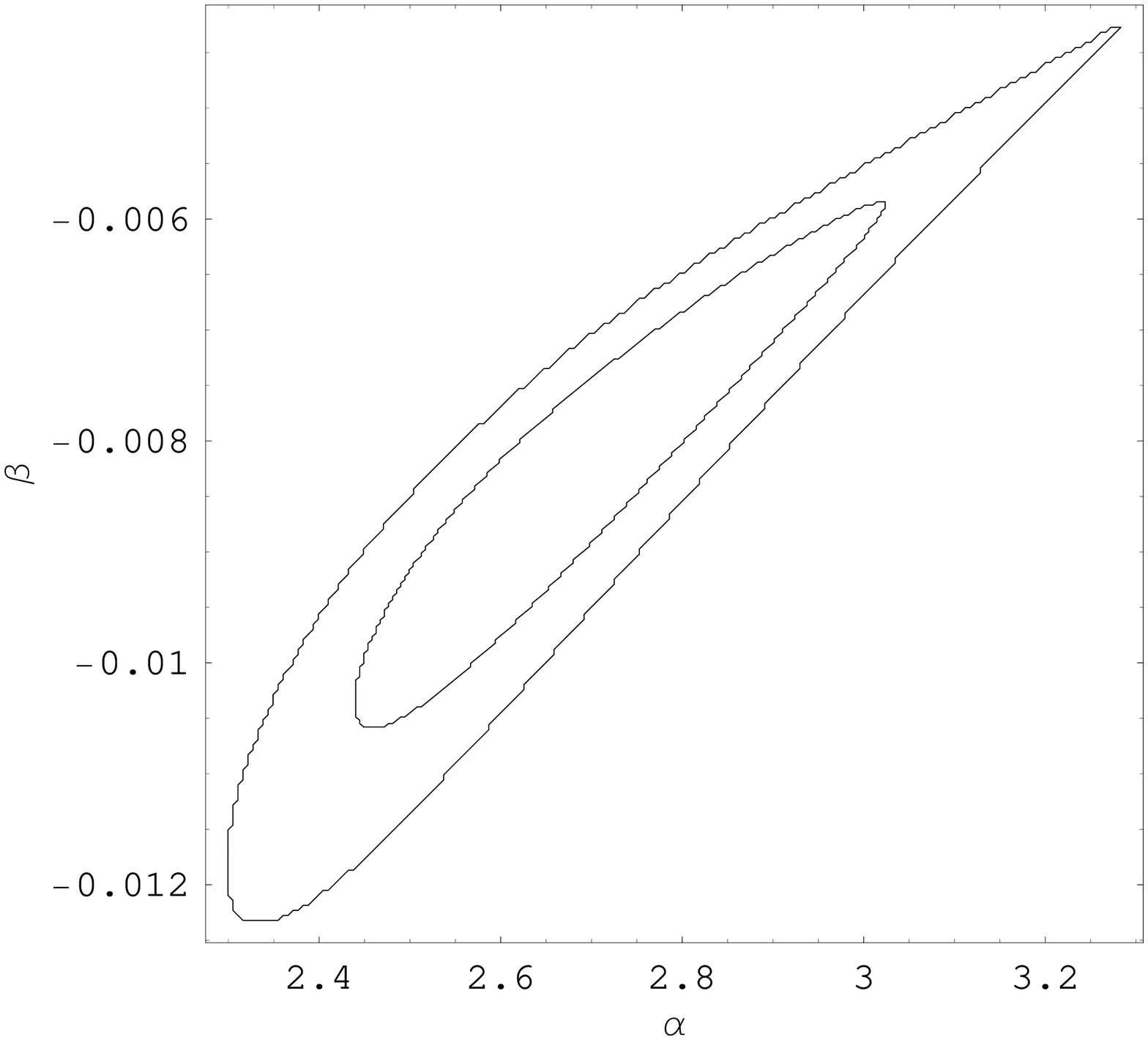}
\caption{The $68.3\%$ and $95.4\%$ confidence levels for the model
 $f(R)=R+\alpha R^2 -\beta/R$ constrained by 
SNLS, BAO and CMB data.} 
\label{forR-R2Rn1and2sig}
\end{figure}

\begin{figure}[h]
\centering
\includegraphics[width=12cm]{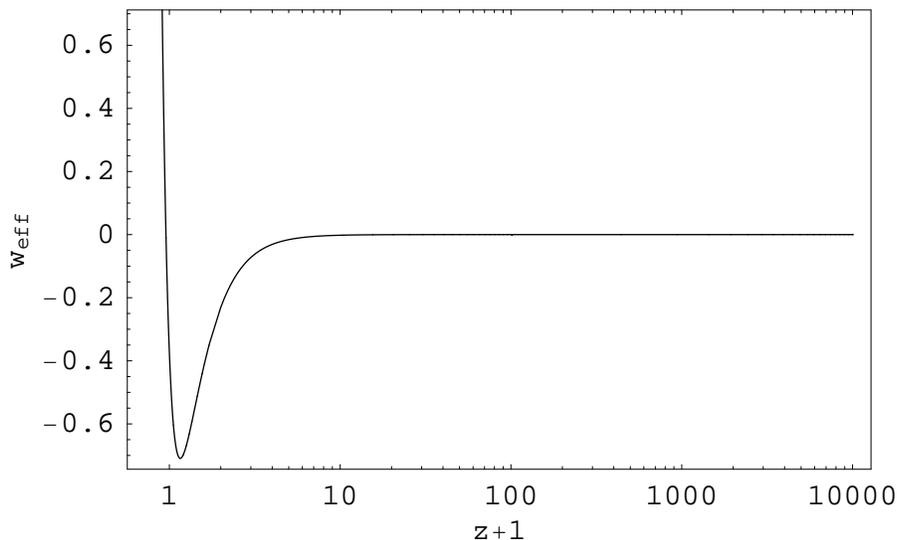}
\caption{\label{fofR-R2RnWeff}
Evolution of $w_{\rm eff}$ for the model 
$R+\alpha R^2-\beta/R$ with the best-fit value 
$\alpha=2.69$ and $\beta=-0.008$.
There is no radiation-dominated epoch prior to the matter-dominated
era ($w_{\rm eff}=0$).
Although $w_{\rm eff}$ can be smaller than $-1/3$ around 
the present epoch, the late-time evolution does not 
correspond to a de-Sitter universe. 
}
\end{figure}

When $m=2$ we also showed separately in Section \ref{m2sin}
that the radiation-dominated epoch does not exist.
To study the observational constraints
in such cases, we have carried out a likelihood analysis 
for the model $f(R)=R+\alpha R^2 -\beta/R$ considered 
in Ref.~\cite{MenWan04}.
Note that the four parameters $\alpha$, $\beta$, $m$ and $n$ 
cannot be constrained at the same time with the observations 
used here, so we have chosen special values for 
$m$ and $n$. With these choices,
we have found that the model is compatible with 
the observational data subject to the parameter constraints
$\alpha\in\left[2.30,3.29\right]$ and 
$\beta\in\left[-0.012,-0.004\right]$ (see Fig.~\ref{forR-R2Rn1and2sig}).
However there is no radiation-dominated stage prior to the 
matter-dominated era, as is seen in Fig.~\ref{fofR-R2RnWeff} which 
gives a plot of the evolution of $w_{\rm eff}$ in the best-fit case
($\alpha=2.69$ and $\beta=-0.008$). 

When $\alpha>0$ and $\beta<0$ we have already shown 
in Section \ref{m2sin} that a singularity appears for $w_{\rm eff}$.
In fact Fig.~\ref{fofR-R2RnWeff} shows that 
the de-Sitter solution is not the late-time attractor since 
there is a singularity in the future (around $z \sim -0.27$) where 
$w_{\rm eff}$ becomes (positive) infinite (and the Hubble parameter 
$H \to 0$) after a short transient period during which 
it is smaller than $-1/3$.

The above discussion demonstrates the necessity of the theoretical 
considerations in the previous section. 
Using the observational data alone,
we would obtain the misleading result that the above model is consistent 
with observations despite the absence of the radiation-dominated era. 
This is due to the fact that the observations we have considered 
do not probe the radiation-dominated phase.

%%%%%%%%%%%%%%%%%%%%%%%%%%%%
%%%%%%%%%%%%%%%%%%%%%%%%%%%%

%%%%%%%%%%%%%%%%%%%%%%%%%%%%
%%%%%%%%%%%%%%%%%%%%%%%%%%%%

%%%%%%%%%%%%%%%%%%%%%%%%%%%%
\begin{table}
\begin{center}
\begin{tabular}{|c||c|c|c|}
\hline
 & $m=0.1$ & $m=0.5$ & $m=0.8$\\
\hline
$n=0.1$ & $(116.68,-1.33,3.37,0.27)$ & $(116.64,-0.22,4.66,0.27)$ & $(116.61,-0.08,4.90,0.27)$\\
$n=0.3$ & $(116.64,-2.47,2.54,0.27)$ & $(116.84,-0.45,5.95,0.27)$ & $(116.76,-0.24,6.15,0.27)$\\
$n=0.5$ & $(116.64,-2.85,2.56,0.27)$ & $(116.63,-0.82,6.36,0.27)$ & -\\
$n=0.8$ & $(116.99,-2.85,4.71,0.28)$ & $(116.76,-1.03,9.16,0.28)$ & $(118.29,-0.47,12.4,0.28)$\\
\hline
\end{tabular}
\caption{The best-fitting values of the parameters
$(\chi^2,\alpha,\beta,\Omega_{m0})$ for the 
models based on theories $f(R)=R+\alpha R^m-\beta/R^n$ 
with some particular values of $(m, n)$.}
\label{tab3}
\end{center}
\end{table}
%%%%%%%%%%%%%%%%%%%%%%%%%%%%

So far we have considered the cases with $m>1$. 
We have also checked that the sequence of radiation-dominated, 
matter-dominated and de-Sitter eras
can be realized for $0<m<1$, 
although in this case the initial inflationary epoch is absent.
One can perform a likelihood analysis by taking several values
of $m$ and $n$ which exist in the regions $0<m<1$ and 
$0<n<1$, while allowing $\alpha$ and $\beta$ to vary. 
Interestingly, there are values of $(m, n)$ in these
ranges which fit the observational data.
Table \ref{tab3} shows the best-fitting model parameters
for some values of the parameters $m$ and $n$.
% in the $(m,n)$ plane. 
For this range of parameters
the best fitting values of 
$\alpha$ and $\beta$ are always found to be negative 
and positive, respectively. 
As $n$ is increased we find that the magnitudes of
$\alpha$ and $\beta$ both increase, while the former
still remaining negative.
The increase in the amplitudes of $\alpha$ and $\beta$ 
tends to compensate for the decrease in the $1/R^n$ term. 
On the other hand as $m$ increases, the amplitude of 
$\alpha$ decreases (while still remaining negative)
while that of $\beta$ increases.
These effects have the consequence of 
minimising the contribution of $R^m$ 
relative to $1/R^n$. We also did not find the divergence 
of $w_{\rm eff}$ in such cases.
Note that from the Table \ref{tab3} the case
with $n=0.8$ and $m=0.8$ can produce late-time acceleration.
This provides an example of a case where the two nonlinear
terms are comparable and necessary to produce the acceleration
at late times.
There would be no late-time acceleration 
if a single term is chosen with such an exponent.

%-------------------------------------------------------------------
\subsection{Theories of the type $f(R)=R+\alpha \ln R-\beta$}
 \label{s33}
%-------------------------------------------------------------------

Finally we consider theories of the type
\begin{eqnarray}
\label{model3}	
f(R) =R+\alpha \ln R-\beta\,,
\end{eqnarray}
where $\alpha$ and $\beta$ are dimensionless constants. 

\begin{figure}
\includegraphics[width=8cm]{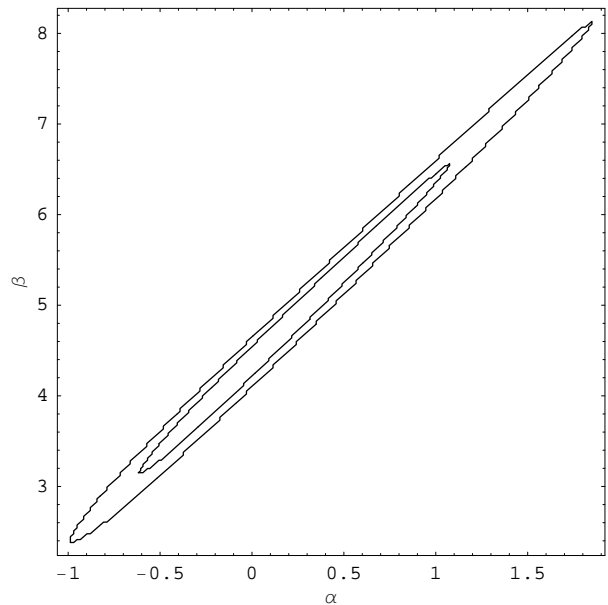}
\caption{$68.3\%$ and $95.4\%$ confidence level for the theory 
$f(R)=R+\alpha \ln R-\beta$ constrained by SNLS, BAO and CMB data.
The models without a cosmological constant $(\beta=0)$
are excluded by the data.
} 
\label{forR-Log}
\end{figure}

The combined constraints from the
SNLS, BAO and CMB data confine the allowed values of $\alpha$
and $\beta$ to the ranges $\alpha\in\left[-1,1.86\right]$ and 
$\beta\in\left[-0.96, 8.16\right]$, respectively (see Fig.~\ref{forR-Log}).
The best-fit model corresponds to  
$\alpha=0.11$ and $\beta=4.62$ with $\chi^2=116.69$.
These values are not too different from the $\Lambda$CDM model
with $\alpha=0$ and $\beta=4.38$, indicating that the
model is not significantly preferred 
over the $\Lambda$CDM model.

Note that the case $\beta=0$ is excluded by the data. 
This therefore shows that the ${\rm ln}\,R$ term alone 
cannot drive the late-time acceleration; one still
requires a cosmological constant. 
Thus based on the observational constraints considered here, 
only strong theoretical reasons would motivate the 
consideration of theories of this type.

%%%%%%%%%%%%%%%%%%%%%%%%%%%%
%%%%%%%%%%%%%%%%%%%%%%%%%%%%

%----------------------------------------------------------------%
\section{Conclusions} \label{s5}
%----------------------------------------------------------------

We have made a detailed and systematic study of a number of 
families of $f(R)$ theories in Palatini formalism, using phase
space analysis, extensive numerical simulations
as well as observational constraints from recent data.
These theories have been recently put forward in the literature in 
order to explain the different dynamical phases in the evolution of 
the universe, in particular the early and/or late acceleration 
epochs deduced from recent observations.
The important advantages of theories based on
Palatini formalism (as opposed to their metric counterparts)
include the facts that they are second
order, have the correct Newtonian limit, pass the solar system tests
and are free from instabilities. This provides important
motivation for their consideration.

Considering the FLRW setting, 
we have expressed the evolution equations
as an autonomous dynamical system in the form of 
Eqs.~(\ref{auto1}) and (\ref{auto2}).
The $\Lambda$CDM model corresponds to a case
where the expression $C(R)$ given by Eq.~(\ref{CR}) vanishes.
Models, based on $f(R)$ theories, in which the variable
$C(R)$ asymptotically approaches $0$ are able to 
give rise to a de-Sitter expansion with constant $R$ at late times.
This includes the models based on theories 
of the type (a) $f(R)=R-\alpha/R^n$, 
(b) $f(R)=R+\alpha R^m-\beta/R^n$,
and (c) $f(R)=R+\alpha {\rm ln}\, R-\beta$.
We have carried out a detailed analysis of the cosmological 
evolution for such models by studying their fixed points and their
stabilities against perturbations.

For models based on theories of the type $f(R)=R-\beta/R^n$,
we have shown that the sequence of radiation-dominated, 
matter-dominated and de-Sitter eras is in fact 
realised for $n>-1$.
This is a stark contrast to the corresponding theories 
of metric formalism for which it is difficult to realize
such a sequence \cite{APT}.

Considering models based on theories of the type 
$f(R)=R+\alpha R^m-\beta/R^n$, we
have found that while they are capable of  
producing early as well as late accelerating phases
an early inflationary epoch does not seem to be
followed by a standard radiation-dominated era.
When $m>2$, the de-Sitter point is stable, which
prohibits an exit from the inflationary phase.
Inflationary solutions can also 
be obtained for $3/2<m<2$ (for $\alpha >0$), but 
in that case they directly exit into a matter-dominated phase
which in turn leads to a late-time accelerating phase.
For $0<m<4/3$, the radiation fixed point is unstable, 
which makes it possible to have the sequence of 
radiation-dominated, matter-dominated and de-Sitter phases
without a preceding inflationary epoch. 
For $4/3<m<3/2$, inflation 
is not possible and radiation, which is a saddle point, 
can only be preceeded by a non-accelerated phase such as $P_{r2}$ or 
an unknown phase not satisfiyng our late and early time assumptions.

We have also placed observational constraints on the parameters 
of these models, employing the recently
released supernovae data by the Supernova Legacy Survey as well as
the baryon acoustic oscillation peak in the SDSS
luminous red galaxy sample and the CMB shift parameter.
We have found that both classes of models are in agreement with the
observations with appropriate choices of their parameters. 
For models based on theories of the type $f(R)=R-\alpha/R^n$, the best
fit values were found to be $\beta=4.63$ and $n=0.027$. 
This is consistent with the results obtained by 
Amarzguioui {\it et al.} \cite{AmaElg06}
who used the supernova `Gold' data
rather than the SNLS without taking into account the radiation.

For the models based on theories of the type
$f(R)=R+\alpha R^m-\beta/R^n$, it is not possible to 
constrain all four parameters simultaneously. 
Concentrating on the special class
of theories $f(R)=R+\alpha R^2-\beta/R$ studied 
in litterature, we have found that they are compatible with the data 
subject to parameter constraints
$\alpha\in\left[2.30,3.29\right]$ and  $\beta\in\left[-0.012,-0.004\right]$.
Despite this agreement we have found, using a phase space analysis, 
that such cases do not seem to produce a radiation-dominted era prior 
to the matter-dominated epoch, thus making them cosmologically 
unacceptable. The reason for this apparent contradiction is that
the combination of the SN, SDSS and the CMB shift parameter
considered here does not provide constraints on early
phases of the universe prior to the decoupling epoch.
We have also checked that a sequence of radiation-dominated, 
matter-dominated and de-Sitter phases becomes possible 
for models based on theories of the type
$f(R)=R+\alpha R^m-\beta/R^n$ when $0<m<1$, 
in agreement with the data.

Finally we have studied the compatibility of theories of
the type $f(R)=R+\alpha {\rm ln}\, R-\beta$ with observations.
Again we have found that such models are compatible 
with observations with
appropriate choices of their parameters. However, the 
logarithmic term on its own is unable to
explain the late-time acceleration consistent with observations.

In summary, we have found that using Palatini formalism
it is possible to produce models based on the classes of 
theories considered here, which possess the sequence of
radiation-dominated, matter-dominated and late-time acceleration
phases consistent with observations. 
However, we have been unable to find the sequence 
of all four phases required for a complete explanation 
of the cosmic dynamics.

%%%%%%%%%%%%%%%%%%%%%%%%%%%%%
\section*{ACKNOWLEDGEMENTS}
S.\,F. is supported by a Marie Curie Intra-European Fellowship of the
European Union (contract number MEIF-CT-2005-515028).
S.\,T. is supported by JSPS
(Grant No.\,30318802).
%%%%%%%%%%%%%%%%%%%%%%%%%%%%%

\section*{Appendix}

In this Appendix we shall give the expression of $w_{\rm eff}$
for the model $f(R)=R+\alpha R^m-\beta/R^n$. We shall 
write the various terms in the expression for $w_{\rm eff}$
in Eq.~(\ref{weff}).

Using the relation (\ref{baeq3}), we obtain
\begin{eqnarray}
\frac{\dot{R}}{HR}
=-3 \frac{1-(m-2)\alpha R^{m-1}-(n+2)\beta R^{-n-1}}
{1-m(m-2) \alpha R^{m-1}+n(n+2) \beta R^{-n-1}}\,,
\end{eqnarray}
which allows the third term in Eq.~(\ref{weff}) to be written as
\begin{eqnarray}
\frac{\dot{F}}{3HF}
=\frac{\dot{R}}{HR} 
\frac{m(m-1)\alpha R^{m-1}-n(n+1)\beta R^{-n-1}}
{3[1+m\alpha R^{m-1}+n \beta R^{-n-1}]}\,.
\end{eqnarray}
Defining the variable $\zeta$, as 
\begin{eqnarray}
\zeta &\equiv& \frac{3F'(FR-2f)}{2F(F' R-F)} \nonumber \\
&=& \frac{3[m(m-1)\alpha R^{m-1}-n(n+1)\beta R^{-n-1}]
[-1+(m-2)\alpha R^{m-1}+(n+2)\beta R^{-n-1}]}
{2(1+m\alpha R^{m-1}+n\beta R^{-n-1})
[-1+m(m-2)\alpha R^{m-1}-n(n+2)\beta R^{-n-1}]}\,,
\end{eqnarray}
allows the fourth term in Eq.~(\ref{weff}) to be expressed as
\begin{eqnarray}
\frac{\dot{\xi}}{3H\xi} &=&-\frac23 \frac{\zeta}{1-\zeta}
\frac{\dot{R}}{HR} \nonumber \\
& & \times\Bigg[ \frac{m(m-1)^2\alpha R^{m-1}+n(n+1)^2\beta R^{-n-1}}
{m(m-1)\alpha R^{m-1}-n(n+1)\beta R^{-n-1}}-
\frac{(m-2)(m-1)\alpha R^{m-1}-(n+1)(n+2)\beta R^{-n-1}}
{1-(m-2)\alpha R^{m-1}-(n+2)\beta R^{-n-1}} \nonumber \\
& &-\frac{m(m-1)\alpha R^{m-1}-n(n+1)\beta R^{-n-1}}
{1+m\alpha R^{m-1}+n\beta R^{-n-1}}+
\frac{m(m-1)(m-2)\alpha R^{m-1}+n(n+1)(n+2)\beta R^{-n-1}}
{1-m(m-2)\alpha R^{m-1}+n(n+2)\beta R^{-n-1}}
\Bigg]\,.
\end{eqnarray}
Finally the last term in Eq.~(\ref{weff}) is given by 
\begin{eqnarray}
\frac{\dot{F}R}{18F \xi H^3}
=\frac{m(m-1)\alpha R^{m-1}-n(n+1)\beta R^{-n-1}}
{3[(m-1)\alpha R^{m-1}+(n+1)\beta R^{-n-1}]}
y_1 \frac{\dot{R}}{HR}\,.
\end{eqnarray}
Note that from Eq.~(\ref{gere}) $R$ is a function of 
$y_1$ and $y_2$, which in turn implies that $w_{\rm eff}$
is also a function of $y_1$ and $y_2$, albeit of a very 
complicated form.

Despite this complexity, however, the expression for $w_{\rm eff}$ 
reduces to a fairly simple forms for specific cases.
For example, when $m=2$ and $\beta=0$, we find 
\begin{eqnarray}
    \label{weffape1}  
w_{\rm eff}=y_{1}+\frac13 y_{2}+
\frac{2\alpha R (2+\alpha R)}{(1+2\alpha R)(1-\alpha R)}\,.
\end{eqnarray}
Now since from Eq.~(\ref{gere}) $\alpha R=2y_1/(1-y_1-y_2)$ 
in this case, $w_{\rm eff}$ expressed in terms of 
$y_1$ and $y_2$ thus
\begin{eqnarray}
\label{weffape2}    
w_{\rm eff}=y_1+\frac13 y_2+\frac{8y_1 (1-y_2)}
{(1+3y_1-y_2)(1-3y_1-y_2)}\,.
\end{eqnarray}
%

%----------------------------------------------------%
\bibliographystyle{unsrt}

\end{document}